\documentclass[aps,prc,twocolumn,showpacs,preprintnumbers,amsmath,amssymb,superscriptaddress]{revtex4-1}

\usepackage{graphicx}% Include figure files
\usepackage{dcolumn}% Align table columns on decimal point
\usepackage{bm}% bold math
\usepackage{epsfig}% Include figure files
\usepackage{multirow}

\begin{document}

\preprint{APS/123-QED}

\title{Investigation of $\alpha$-induced reactions on $^{127}$I for the astrophysical $\gamma$ process}

\author{G.\,G.\,Kiss}%
 \email{ggkiss@atomki.hu}
\affiliation{%
Institute of Nuclear Research (ATOMKI), H-4001 Debrecen, POB.51., Hungary}%
\author{T.\,Sz\"ucs}%
\affiliation{%
Institute of Nuclear Research (ATOMKI), H-4001 Debrecen, POB.51., Hungary}%
\author{Zs.\,T\"or\"ok}%
\affiliation{%
Institute of Nuclear Research (ATOMKI), H-4001 Debrecen, POB.51., Hungary}%
\author{Z.\,Korkulu}%
\affiliation{%
Institute of Nuclear Research (ATOMKI), H-4001 Debrecen, POB.51., Hungary}%
\affiliation{%
Kocaeli University, Department of Physics, TR-41380 Umuttepe, Kocaeli, Turkey}
\author{Gy.\,Gy\"urky}%
\affiliation{%
Institute of Nuclear Research (ATOMKI), H-4001 Debrecen, POB.51., Hungary}%
\author{Z.\,Hal\'asz}%
\affiliation{%
Institute of Nuclear Research (ATOMKI), H-4001 Debrecen, POB.51., Hungary}%
\author{Zs.\,F\"ul\"op}%
\affiliation{%
Institute of Nuclear Research (ATOMKI), H-4001 Debrecen, POB.51., Hungary}%
\author{E.\,Somorjai}%
\affiliation{%
Institute of Nuclear Research (ATOMKI), H-4001 Debrecen, POB.51., Hungary}%
\author{T.\,Rauscher}%
\affiliation{%
Department of Physics, University of Basel, CH-4056 Basel, Switzerland}%

\date{\today}

\begin{abstract}

\textbf{BACKGROUND}: The $\gamma$ process in core-collapse supernova explosions
is thought to explain the origin of proton-rich isotopes between Se and
Hg, the so-called $p$ nuclei. The majority of the reaction rates for
$\gamma$-process reaction network studies has to be predicted in
Hauser-Feshbach statistical model calculations. Recent investigations
showed problems in the prediction of $\alpha$ widths at astrophysical
energies. This impacts the reliability of abundance predictions in the
upper mass range of the $p$ nuclei. 

\textbf{PURPOSE}: Measurement of the $^{127}$I($\alpha$,$\gamma$)$^{131}$I and
$^{127}$I($\alpha$,n)$^{130}$I reaction cross sections close to the
astrophysically relevant energy range to test the predictions, to derive
an improved reaction rate, and to extend the database required to define
an improved global optical $\alpha$+nucleus potential. 

\textbf{METHODS}: The cross sections were derived using the activation technique, the yield of the emitted $\gamma$ and characteristic X-ray photons were measured using a LEPS and a HPGe detector.

\textbf{RESULTS}: The cross sections of the $^{127}$I($\alpha$,$\gamma$)$^{131}$Cs reaction have been determined for the first time, at energies $9.50\leq E_{c.m.} \leq 15.15$ MeV. The $^{127}$I($\alpha$,n)$^{130}$Cs reaction was studied in the range  $9.62\leq E_{c.m.} \leq 15.15$ MeV.
Furthermore, the relative intensity of the 536.1 keV $\gamma$ transition was measured precisely, its uncertainty was reduced from 13\% to 4\%. 
The results were then compared to Hauser-Feshbach
calculations which were also used to extend the cross sections into the
astrophysically relevant region and to compute the reaction rate. 

\textbf{CONCLUSIONS}: The comparison to statistical Hauser-Feshbach model
calculations showed that the $\alpha$ width can be described well in the
measured energy range using a standard, energy-independent global
optical potential. The newly derived stellar reaction rates at
$\gamma$-process temperatures for $^{127}$I($\alpha$,$\gamma$)$^{131}$I
and its reverse reactions, nevertheless, are faster by factors $4-10$
than those from previous calculations, due to further improvements in
the reaction model. The importance of the inclusion of complete level
schemes into the Hauser-Feshbach calculations is illustrated by
comparing the impact of two different level schemes, one of them
extending to higher excitation energy but not containing all relevant
levels. 

\end{abstract}


\maketitle

\section{Astrophysical motivation}
\label{sec:intro}

The bulk of the stable isotopes heavier than iron observed in our Solar System are produced via neutron capture reactions \cite{b2fh}. In order to reproduce the abundance pattern of the heavy isotopes, two different neutron capture processes have to be assumed: the \textit{s}-process (which has a main \cite{gal98} and a weak component \cite{rau02}) and the \textit{r}-process \cite{ctt,arn07}.

On the proton-rich side of the valley of stability, however, there are 35 nuclei which are separated from the path of the neutron capture processes. These mostly even-even isotopes between $^{74}$Se and $^{196}$Hg are the so-called \textit{p} nuclei \cite{woo78}. It is generally accepted that the main stellar mechanism synthesizing the \textit{p} nuclei -- the so-called $\gamma$ process -- involves mainly photodisintegrations, dominantly ($\gamma$,n) reactions on preexisting, neutron-richer \textit{s} and \textit{r} seed nuclei. The high energy photons -- necessary for the $\gamma$-induced reactions -- are available in explosive nucleosynthetic scenarios like a core-collapse supernova shockwave moving through the Ne/O layer of a massive star and reaching temperatures around a few GK. Favorable conditions have recently been found also in Type Ia Supernovae \cite{travaglio}.

Consecutive ($\gamma$,n) reactions can drive the material towards the proton rich side of the valley of stability. As the neutron separation energy increases along this path, ($\gamma$,p) and ($\gamma$,$\alpha$) reactions become stronger and bend the reaction flow towards lighter elements \cite{arn03,rau06,rap06}. Theoretical investigations agree that ($\gamma$,p) reactions are more important for the lighter \textit{p} nuclei while ($\gamma$,$\alpha$) reactions are mainly important at higher masses (A $\geq$ 100). However, calculations based solely on the $\gamma$ process in massive stars are unable to reproduce the observed abundance pattern of the \textit{p} nuclei. 
To solve this problem, several processes have been suggested to fill in the missing nuclear abundances, e.g., the rapid proton capture (\textit{rp}) process on the surface of accreting neutron stars \cite{sch98,sch01} and the neutrino-induced \textit{p} process ($\nu p$ process) in the deepest layers of a core-collapse supernova ejected in a neutrino-wind \cite{fro06,fro12}. Considerations regarding isotope ratios in meteoritic material, however, severely constrain their possible contribution \cite{dauphas}.

Theoretical studies of the nuclear uncertainties in the nucleosynthesis of the \textit{p} nuclei
make use of large reaction networks with mainly theoretical reaction rates (taken from the Hauser-Feshbach (H-F) model \cite{hf}). They have shown that the reaction flow for the production of heavy \textit{p} nuclei (140 $\leq$ A $\leq$ 200) is strongly sensitive to the ($\gamma,\alpha$) photodisintegration rates \cite{rau06,rap06}.
Recent experiments, however, indicate that the H-F predictions may overestimate the $\alpha$-capture cross sections at low energies by factors $3-10$. This would also strongly impact the astrophysical reaction rates. Unfortunately, the available data are scarce, only a handful of reactions have been studied at sufficiently low energies \cite{ful96,rap01,gyu06,ozk07,cat08,yal09,som98,gyu10,kiss_plb,sauer11}.
Consequently, experimental data is urgently needed to confirm the path of the \textit{$\gamma$} process, especially for reactions involving $\alpha$ particles.

It has to be understood that straightforward measurements of ($\gamma$,$\alpha$) reactions on target nuclei in the ground state (g.s.) cannot provide the required information for astrophysical ($\gamma$,$\alpha$) reactions in a stellar plasma, involving a large number of reactions on nuclei in thermally excited states. The sensitivity of the photodisintegrations on g.s.\ nuclei is very different from the one of stellar rates and measurements, therefore test the prediction of different nuclear properties than actually entering the calculation of the rates \cite{raureview,sensi}. In a high-temperature stellar plasma, as encountered in the $\gamma$ process, there is complete detailed balance between forward and reverse rates, therefore ($\gamma,\alpha$) rates can be easily converted to ($\alpha,\gamma$) ones. This is only possible with \textit{stellar} rates, however \cite{raureview,fow74,hwfz}. If the g.s.\ contribution to the stellar rate $0\leq X\leq 1$ is close to unity, i.e., reactions on the g.s.\ of the target nucleus provide most of the rate, a measurement will be able to directly provide the rate \cite{sensi,xfactor}. Photodisintegrations have several orders of magnitude lower $X$ than captures \cite{sensi,moh07}. Therefore, a measurement of ($\alpha,\gamma$) is required to test the predictions of stellar ($\gamma,\alpha$) rates.

It is essential, however, to measure as close as possible to the astrophysically relevant energies, i.e., the energies from which the largest contribution to the reaction rate integral are coming \cite{raureview,tommy_gamow}. Because of the strong energy dependence of charged-particle reaction widths, the dependences of the reaction cross sections at astrophysical, subCoulomb energy and those at higher, measured energy can be quite different. For example, all astrophysical $\alpha$ captures are determined by the $\alpha$ width, whereas well above the Coulomb barrier the $\gamma$ width is mostly dominating and other particle widths may play a role close to channel openings \cite{raureview,sensi}. If a measurement is not possible in the $\alpha$ dominated energy region, the simultaneous measurement of a reaction channel dominated by the $\alpha$ width may prove useful, e.g., the ($\alpha$,n) channel. The problem with this is that the ($\alpha$,n) reaction on proton-rich target nuclei, interesting for the $\gamma$ process, always exhibits a negative $Q$ value, thus limiting the measurable energy range again to higher energies than astrophysically relevant. Nevertheless, this can be used as an initial test of the predicted $\alpha$ width. As further discussed in Sec.~\ref{sec:res}, it remains important to remember that a discrepancy between theory and experiment in $\alpha$ capture does not necessarily imply an incorrect prediction of ($\alpha,\gamma$) and ($\gamma$,$\alpha$) reaction rates in stellar plasmas.

In order to extend the experimental database for the astrophysical $\gamma$ process and to test the reliability of H-F statistical model predictions
in the A $\geq$ 100 mass range, the cross sections of the $^{127}$I($\alpha$,n) and $^{127}$I($\alpha$,$\gamma$) reactions have been measured for the first time, using the activation method. The $^{127}$I($\alpha$,n) reaction was studied in the range $9.62 \leq E_{c.m.} \leq 15.15$ MeV. The radiative $\alpha$-capture cross sections were measured at $9.50 \leq E_{c.m.} \leq 15.15$ MeV. This is close to the astrophysically relevant energy region (the Gamow window), which covers $5.8-8.2$ MeV at a plasma temperature of $T = 2.5$ GK \cite{tommy_gamow}.

The paper is organized as follows:
The experimental details are described in Sec.~\ref{sec:exp}; the resulting cross sections are presented in Sec.~\ref{sec:res}, where also a comparison to predictions is shown and the implications for the astrophysical rate are discussed;
Sec.~\ref{sec:sum} provides the conclusions and a short summary.

\section{Experimental technique}
\label{sec:exp}

\begin{table*}
\caption{\label{tab:decay}Decay parameters of the $^{130,131}$Cs reaction products, formed via the $^{127}$I($\alpha$,n)$^{130}$Cs and the $^{127}$I($\alpha$,$\gamma$)$^{131}$Cs reactions. The relatively weak characteristic K$_{\beta}$ X-ray transitions (indicated by italics) were only used to measure the relative intensity of the 536.1 keV $\gamma$ transition.}
\setlength{\extrarowheight}{0.1cm}
\begin{ruledtabular}
\begin{tabular}{ccccccc}
\parbox[t]{0.8cm}{\centering{Product \\ nucleus}} &
\parbox[t]{1.2cm}{\centering{Decay mode}} &
\parbox[t]{1.2cm}{\centering{Half-life (hour)}} &
\parbox[t]{2.0cm}{\centering{transition}} &
\parbox[t]{2.0cm}{\centering{X- and $\gamma$-ray \\energy (keV)}} &
\parbox[t]{2.0cm}{\centering{Relative $\gamma$-intensity \\ per decay (\%)}} &
\parbox[t]{1.cm}{\centering{Ref.}} \\
\hline
$^{131}$Cs & $\epsilon$ 100\% &  232.54 $\pm$ 0.38   & K$_{\alpha2}$ &29.461 & 21.1 $\pm$ 0.5 & \cite{NNDC1} \\
& &                                                  & K$_{\alpha1}$ &29.782 & 38.9 $\pm$ 0.9  & \\
$^{130}$Cs & $\epsilon$ 96.4\%& 0.4868 $\pm$ 0.00067 & K$_{\alpha2}$ &29.461 & 21.6 $\pm$ 0.7  & \cite{NNDC2}\\
& &																									 & K$_{\alpha1}$ &29.782 & 11.7 $\pm$ 0.6  & \\
& &																									 & \textit{K$_{\beta3}$}  &\textit{33.56}  & \textit{2.05} $\pm$ \textit{0.05} & \\
& &																									 & \textit{K$_{\beta1}$}  &\textit{33.62}  & \textit{3.95} $\pm$ \textit{0.10} & \\
& &																									 & \textit{K$_{\beta2}$}  &\textit{34.42}  & \textit{1.20} $\pm$ \textit{0.03} & \\
& & & & 536.1 & 3.80 $\pm$ 0.50 &  \\
& & & &       & 3.81 $\pm$ 0.15 & present work \\
\end{tabular} \label{decaypar}
\end{ruledtabular}
\end{table*}

The element iodine has only one stable isotope $^{127}$I. The ($\alpha$,n) and ($\alpha,\gamma$) reactions studied in the present work lead to unstable cesium ($^{130,131}$Cs) isotopes. The $^{130}$Cs reaction product has a half-life of 29.21 minutes and its electron capture decay is followed by the emission of a 536.1 keV $\gamma$ ray. The half-life of the $^{131}$Cs produced via the $^{127}$I($\alpha,\gamma$) reaction is 9.69 days. It decays exclusively by electron capture and no $\gamma$ radiation is emitted during its decay. This electron capture is followed by the emission of characteristic X-rays. For the determination of the alpha-capture cross sections the yield of these characteristic X-rays was used similarly as in \cite{kiss_plb}. The technical details of this approach are described in \cite{kiss_npa}.
Further experimental details are described below in details. The decay properties of the reaction products are summarized in Table \ref{tab:decay}.

\subsection{Target production and irradiation}
\label{sec:pro}

Because of its high electronegativity, iodine reacts violently with alkali metals such as Na and K. Therefore, it naturally occurs mostly in compounds. For the target production a KI compound was used because it is chemically stable and its melting point is sufficiently high.
The targets were made by vacuum evaporation from a Mo boat onto 2 $\mu$m thick aluminum foils. During the evaporation, the distance between the Al backing and the evaporation boat was 9 cm. Using this relatively large distance, uniform targets could be produced.

The absolute number of target atoms was determined by PIXE (Proton Induced X-ray Emission) method \cite{pixe} at the PIXE chamber installed on the left 45$^{\circ}$ 
beamline of the 5 MV VdG accelerator of ATOMKI, where the current of the beam and the dead time can be measured very precisely. A more detailed description 
of the setup can be found in \cite{pixe2}. A homogeneous beam of 2 MeV protons with 0.5 mm diameter and $\approx$ 1 nA current was used for the thickness measurement. The total collected charge in the case of each target was about 1 $\mu$C. A typical PIXE spectrum is shown in Figure 1. The spectrum was fitted using the PIXEKLM program code \cite{pixe3}. The peaks used for the analysis are marked. 

The thickness of the KI targets were between 106 and 507 $\mu$g/cm$^2$, corresponding to about 5.3 $\times$ 10$^{17}$ I atoms/cm$^2$ and 24.0 $\times$ 10$^{17}$ I atoms/cm$^2$. The X-ray attenuation factor for the K$_{\alpha1,2}$ characteristic X-rays emitted during the decay of the alpha-capture reaction products was calculated using the LISE code \cite{LISE} and was found to be below 0.4\% for these target thicknesses. Since the chemical composition of the target is important for the energy-loss calculations, the ratio of the K and I atoms in the target were also determined and were found to be 1:1 (within 1\%). 
The precision of the determination of the number of target atoms was better than 4\,\%. Using the PIXE method, the following impurities have been found in the target and the backing: Fe, Cu, Ga, Zn, V (less than 50 ppm each). To check the PIXE results, weighing was also used to determine the
number of target atoms in a few cases. The agreement between the results of the two methods is within 5\%.

\begin{figure}
\includegraphics[angle=0,width=\columnwidth]{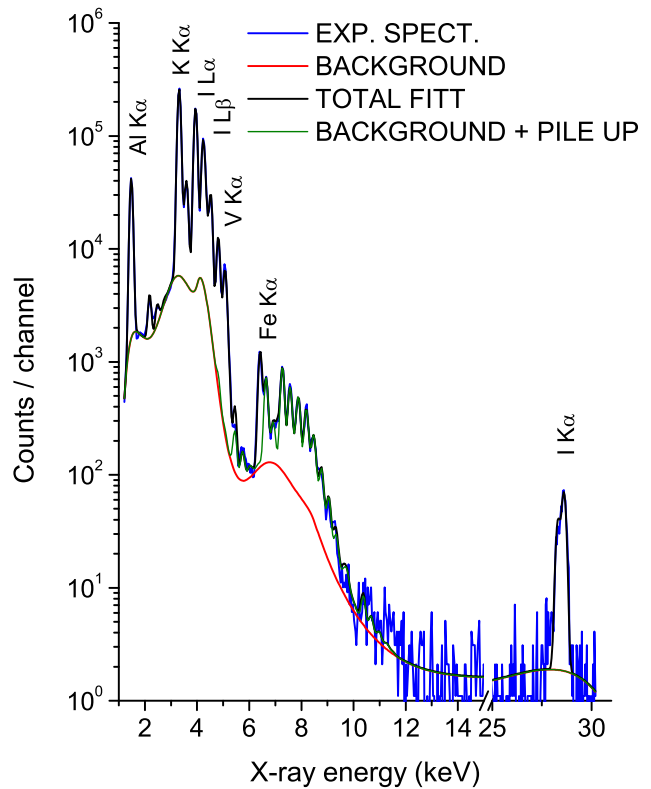}
\caption{\label{fig:pixe} (Color online) PIXE spectrum measured by bombarding the KI targets with 2 MeV protons. The peaks used for the analysis
are marked. Peaks belonging to impurities in the target and/or the backing are indicated, too.}
\end{figure}

The KI targets have been irradiated with $\alpha$ beams from the MGC20 cyclotron of ATOMKI. The energies were in the range of $9.81\leq E_\alpha \leq 15.5$ MeV, covered in steps of about 0.5 MeV. After the beam-defining aperture, the chamber was insulated and a secondary electron suppression voltage of $-300$ V was applied at the entrance of the chamber \cite{yal09}. The typical beam current was between 0.1 and 0.6 p$\mu$A, the length of each irradiation was between 0.92 and 24 hours, corresponding to about 2.7 $\times$ 10$^{15}$ and 2.7 $\times$ 10$^{17}$ total incident $\alpha$ particles. Several beam tests were performed to check the target stability before the experiment. These tests showed that there was no deterioration of the targets using a beam current less than 0.8\,p$\mu$A. The current integrator counts were recorded in multichannel scaling mode, stepping the channel every minute to take into account the changes in the beam current. 

\begin{figure}
\includegraphics[angle=0,width=\columnwidth]{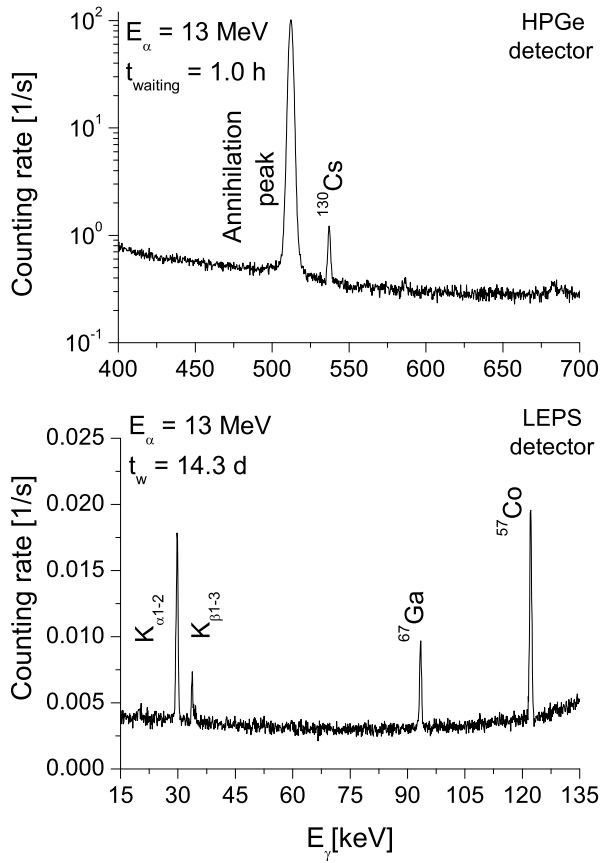}
\caption{\label{fig:spectra} Off-line $\gamma$ - (upper panel) and characteristic X-ray (lower panel) spectra normalized to the length of the countings, taken after irradiating a $^{127}$I target with E$_{\alpha}$ = 13.0 MeV(E$_{c.m.}$ = 12.58 MeV)
beam. The $\gamma$-line and X-ray transitions used to measure the cross section of the $^{127}$I($\alpha$,n)$^{130}$Cs and $^{127}$I($\alpha,\gamma$)$^{131}$Cs reaction cross sections as well as the main background lines are marked.}
\end{figure}

\subsection{$\gamma$- and characteristic X-ray countings}
\label{sec:countings}

For the $\gamma$ counting a 100\% relative efficiency HPGe detector placed in ultra low background (ULB) configuration was used. The X-ray countings were carried out using a Low Energy Photon Spectrometer (LEPS). In order to reduce the laboratory background a multilayer quasi 4$\pi$ shield has been built around this detector including an inner 4 mm thick layer of copper and a 2 mm thick layer of cadmium and an 8 cm thick outer lead shield \cite{kiss_cgs}. The low yields measured in the present work necessitated the use of small source-to-detector distances both for the $\gamma$- and X-ray countings. The distance between the target and the endcup of the ULB detector was 1 cm. In the case of the X-ray countings, the targets were placed 3 cm from the Be entrance window of the LEPS detector.

The detector efficiencies for both detectors used in the present work had to be known in close geometry with high precision.
A similar procedure had been used to derive the photopeak efficiencies in the case of the HPGe and LEPS detectors. At first the absolute detector efficiency was measured in far geometry: at 10 cm distance from the surface of the detector, using calibrated $^{57}$Co, $^{60}$Co, $^{133}$Ba, $^{137}$Cs, $^{152}$Eu, and $^{241}$Am sources with the HPGe detector, and $^{133}$Ba, $^{152}$Eu, and $^{241}$Am sources with the LEPS detector. Since the calibration sources (especially $^{133}$Ba, $^{152}$Eu) emit multiple $\gamma$-radiations from cascade transitions, in the close geometry a strong true coincidence summing effect is expected resulting in
an increased uncertainty of the measured efficiency. Therefore, no direct efficiency measurement
in close geometry has been carried out. Instead, the activity of several irradiated $^{127}$I targets has been measured both in close
and far geometry. Taking into account the time elapsed between the two countings, a conversion
factor of the efficiencies between the two geometries could be determined and used henceforward in the analysis.

\subsubsection{Measurement of the $^{127}$I($\alpha$,n)$^{130}$Cs reaction products}

The relative intensity of the 536.1 keV $\gamma$ transition - emitted during the decay of the $^{130}$Cs reaction product - is known with limited precision, its uncertainty is 13\%. In order to increase the precision of the cross section data derived in the present work, first, the uncertainty of the branching ratio of this transition was reduced. Two relatively thin KI samples have been irradiated with a $E_{\alpha}$ = 15 MeV ($E_{c.m.}$ = 14.52 MeV) beam and their spectra were measured using the LEPS detector.
The yield of the 536.1 keV $\gamma$ transition was normalized to the yield of the K$_{\alpha1,2}$, to the yield of the K$_{\alpha1,3}$ and to the yield of the K$_{\beta2}$ X-ray transitions, taking into account their well known relative intensities and the detector efficiencies.
The weighed average of the 6 normalized relative intensities was found to be 3.81\% $\pm$ 0.15\%. Its uncertainty was calculated by accounting for the following partial uncertainties: uncertainty of the detector efficiency ratios (3\%), uncertainty of the X-ray relative intensities taken from the literature ($\leq$ 0.5\%) and statistical uncertainty ($\leq$ 2.0\%).

\begin{table}
\caption{\label{tab:results_an} Measured cross sections of the $^{127}$I($\alpha$,n)$^{130}$Cs reaction.}
\begin{ruledtabular}
\begin{tabular}{cc}
$E_\mathrm{c.m.}$ &
cross section \\
(MeV) &
(mbarn) \\
\hline
\multirow{18}{*}{
\begin{tabular}{r@{\,$\pm$\,}l} 
9.62  & 0.10 \\
10.01 & 0.09 \\
10.64	& 0.06 \\
10.71	& 0.08 \\
11.10 & 0.09 \\
11.13	& 0.06 \\
11.58 & 0.06 \\
11.60	& 0.06 \\
12.18 & 0.09 \\
12.22	& 0.06 \\
12.58 & 0.07 \\
12.59	& 0.09 \\
13.18	& 0.09 \\
13.55	& 0.07 \\
14.19	& 0.09 \\
14.52	& 0.07 \\
14.53	& 0.08 \\
15.15	& 0.10	  
\end{tabular}
} &
\multirow{18}{*}{
\begin{tabular}{r@{\,$\pm$\,}l}
0.013 & 0.002 \\
0.045	& 0.007	\\
0.17	& 0.01  \\
0.18	& 0.05  \\
0.59  & 0.05  \\
0.66	& 0.11  \\
1.61  & 0.12  \\
1.53	& 0.27  \\
4.30  & 0.3   \\
5.01	& 0.82  \\
8.78  & 0.61  \\
10.2	& 1.6   \\
17.1	& 2.5 	\\
32.1	& 4.8 	\\
68.6	& 10.1  \\
119.9	& 8.4 	\\
115.6	& 17.0  \\
167.6	& 24.7  	
\end{tabular}
} 
\\ & \\ & \\ & \\ & \\ & \\ & \\ & \\ & \\ & \\ & \\ & \\ & \\ & \\ & \\ & \\ & \\ &
\end{tabular}
\end{ruledtabular}
\end{table}

The activity of the $^{127}$I($\alpha$,n) reaction products was measured after the end of the irradiation by counting the yield of the 536.1 keV $\gamma$ transition at and above $E_\mathrm{c.m.}$ = 10.71 MeV. The upper part of Fig.~\ref{fig:spectra} shows a typical spectrum collected using the HPGe detector after the irradiation of a KI target with 13 MeV $\alpha$ beam ($E_\mathrm{c.m.}$ = 12.58 MeV). At the low energy irradiations (below $E_\mathrm{c.m.}$ = 10.71), the yield of the 536.1 keV transition was not sufficient for the analysis. For this reason, the emitted K$_{\alpha1,2}$ characteristic X-ray photons were counted. To determine the cross sections at these energies, the length of the irradiations was short (about 1 hour), in order to limit the beam-induced background. To verify this approach, several irradiations at higher energies were repeated, the $^{127}$I($\alpha$,n) reaction cross sections were measured two times, first by counting the yield of the 536.1 keV transition, and in another irradiation by measuring the yield of the emitted characteristic K$_{\alpha1,2}$ X-ray photons. The experimental cross sections are given in Table \ref{tab:results_an}.
Between the end of the irradiation and the $\gamma$ / X-ray counting, a waiting time of about 0.25 hours was inserted in order to decrease the yield of the disturbing short-lived activities. All spectra were taken for two hours at most and stored regularly in order to follow the decay of the $^{130}$Cs reaction product.

\subsubsection{Measurement of the $^{127}$I($\alpha,\gamma$)$^{131}$Cs reaction products}
 
The yield of the characteristic X-rays --- emitted during the electron-capture decay of the produced unstable cesium isotopes --- were used for the determination of the ($\alpha,\gamma$) and ($\alpha$,n) reaction cross sections. The disadvantage of this approach is that it is not able to distinguish between the decay of the different isotopes of the same element.
However, using a similar approach as in \cite{kiss_plb}, the two open reaction channels can be separated.
Calculations show that close above the threshold the ($\alpha$,n) channel becomes dominant, its cross section exceeds by orders of magnitude that of the ($\alpha,\gamma$) channel. Since the half-life of $^{131}$Cs is about 478 times longer than that of $^{130}$Cs, if the X-ray countings are carried out more than 1 day after the end of the irradiation, the target's $^{130}$Cs activity decreases to a negligible level and the decay of the longer-lived $^{131}$Cs can be measured.

The energies of the emitted characteristic X-ray K$_{\alpha1,2}$ lines are 29.782 and 29.461 keV, respectively. Since the resolution of the LEPS detector is typically between 400 eV (for a 5.9 keV $\gamma$ line) and 680 eV (for a 122 keV $\gamma$ ray), in the X-ray spectra it is not possible to distinguish between K$_{\alpha1}$ and K$_{\alpha2}$ transitions. Instead, the sum of the emitted characteristic X-rays was used for the analysis. The X-ray spectra were taken for 1-6 days and stored regularly in order to follow the decay of the $^{131}$Cs reaction products. In most of the cases, the X-ray counting was carried out twice and consistent results were found. The lower part of Fig.~\ref{fig:spectra} shows a typical spectrum collected using the LEPS detector after the irradiation of a KI target with a $E_{\alpha}$ = 13 MeV ($E_\mathrm{c.m.}$ = 12.58 MeV) beam.

The experimental cross sections for this reaction channel are given in Table \ref{tab:results_ag}.

\begin{table}
\caption{\label{tab:results_ag} Measured cross sections of the $^{127}$I($\alpha,\gamma$)$^{131}$Cs reaction.}
\begin{ruledtabular}
\begin{tabular}{cc}
$E_\mathrm{c.m.}$ &
cross section \\
(MeV) &
(mbarn) \\
\hline
\multirow{12}{*}{
\begin{tabular}{r@{\,$\pm$\,}l} 
 9.50 & 0.12 \\
10.00 & 0.09 \\
10.71	& 0.08 \\
11.13	& 0.06 \\
11.60 & 0.06 \\
12.22	& 0.06 \\
12.59 & 0.09 \\
13.18	& 0.09 \\
13.55 & 0.07 \\
14.19	& 0.09 \\
14.53 & 0.08 \\
15.15	& 0.10	  
\end{tabular}
} &
\multirow{12}{*}{
\begin{tabular}{r@{\,$\pm$\,}l}
0.18  & 0.04 \\
0.65	& 0.12 \\
1.76	& 0.29 \\
3.90	& 0.79 \\
8.54  & 0.74 \\
21.6	& 1.7  \\
33.4  & 2.6  \\
42.8	& 3.5  \\
63.5  & 5.0  \\
96.6	& 7.3  \\
123.2 & 9.2 \\ 
127.7 & 11.0
\end{tabular}
} 
\\ & \\ & \\ & \\ & \\ & \\ & \\ & \\ & \\ & \\ & \\ &
\end{tabular}
\end{ruledtabular}
\end{table}

\section{Results and discussion}
\label{sec:res}

The measured $\alpha$-induced cross section values are listed in Tables \ref{tab:results_an} and \ref{tab:results_ag}. The quoted uncertainty in the $E_\mathrm{c.m.}$ values corresponds to the energy stability of the $\alpha$-beam and to the uncertainty of the energy loss in the target, which was calculated using the SRIM code \cite{SRIM}. Some irradiations were repeated at the
same energies. The cross sections are then derived from the averaged results of the irradiations weighted by the statistical uncertainty of the measured values. The uncertainty of the cross sections is the quadratic sum of the following partial errors: efficiency of the HPGe detector and LEPS (6 and 5 \%, respectively), number of target atoms (5\%), current measurement (3\%), uncertainty of decay parameters ($\leq$\ 13.2 \%) and counting statistics (1.6 - 17.9\%).

\subsection{Astrophysical implications}

\begin{figure*}
\includegraphics[angle=270,width=\textwidth]{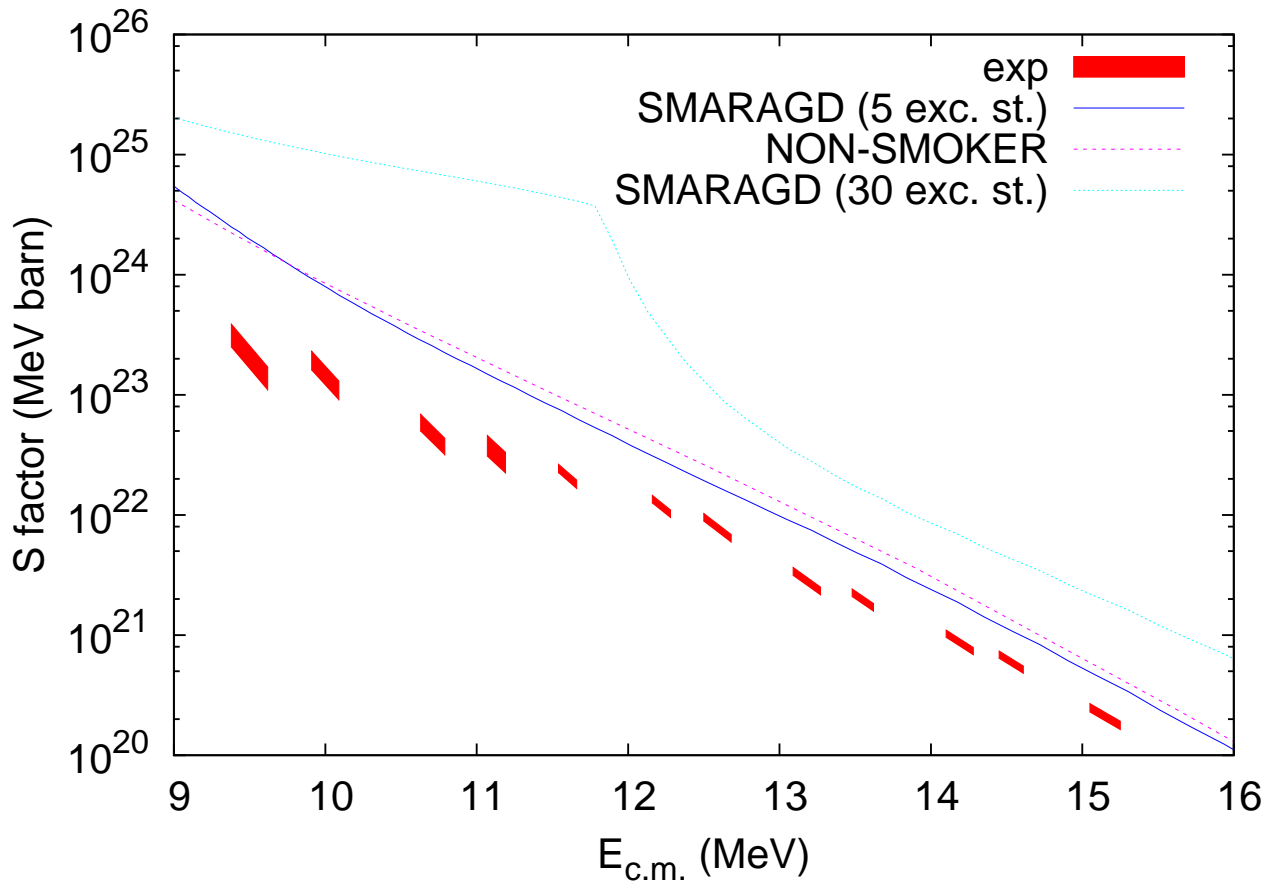}
\caption{\label{fig:ag} (Color online) Experimental $S$ factors of the $^{127}$I($\alpha$,$\gamma$)$^{131}$Cs reaction compared to theoretical predictions with the NON-SMOKER and SMARAGD codes (see text).}
\end{figure*}

\begin{figure*}
\includegraphics[angle=270,width=\textwidth]{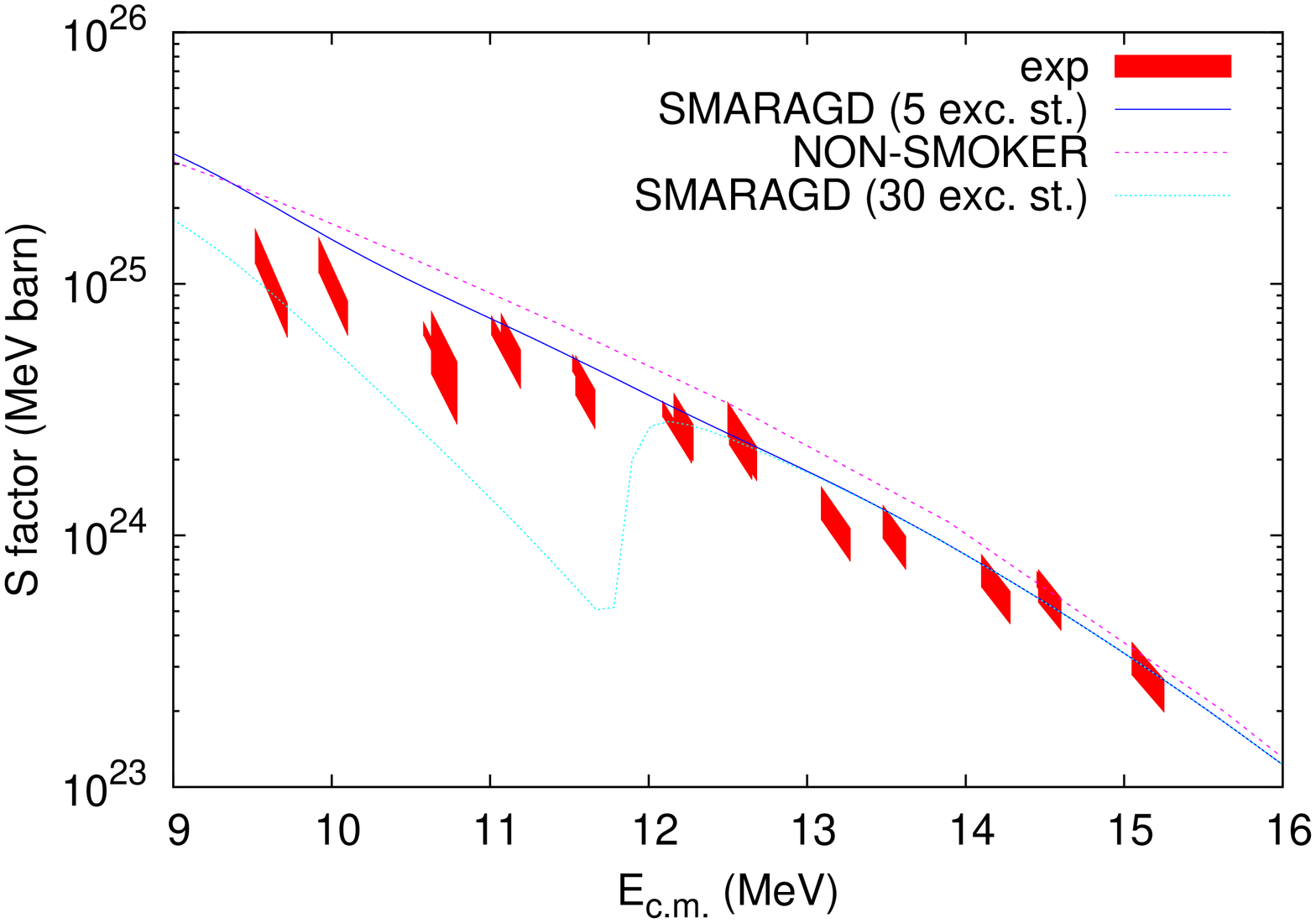}
\caption{\label{fig:an} (Color online) Experimental $S$ factors of the $^{127}$I($\alpha$,n)$^{130}$Cs reaction compared to theoretical predictions with the NON-SMOKER and SMARAGD codes (see text).}
\end{figure*}

In astrophysical investigations it is common to quote the astrophysical $S$ factor $S(E)$ which removes the energy dependence due to the projectile penetration through the Coulomb barrier from the cross sections $\sigma(E)$ \cite{ilibook},
\begin{equation}
 S(E)=E \sigma(E) e^{2\pi\eta}\quad,
\end{equation}
with $\eta$ being the Sommerfeld parameter
\begin{equation}
\eta=\frac{Z_\mathrm{p}Z_\mathrm{T}e^2}{\hbar}\left( \frac{\mu}{2E} \right)^{1/2} \quad.
\end{equation}
The charge numbers $Z_\mathrm{p}$, $Z_\mathrm{T}$ of projectile and target, respectively, and their reduced mass $\mu$ enter the Sommerfeld parameter. When there is an uncertainty in the energy $E$, the calculation of the $S$ factor errors becomes more complicated because
the energy enters also via the Sommerfeld parameter. The error bars on $\alpha$ energy and cross section translate into an error region for the $S$ factor which is of trapezoid shape, instead of the usual rectangular shape. These error trapezoids are also used in Figs.~\ref{fig:ag} and \ref{fig:an}, showing a comparison of the experimental $S$ factors for the ($\alpha$,$\gamma$) and ($\alpha$,n) reactions, respectively, with theoretical calculations in the H-F approach.

\begin{figure}
\includegraphics[angle=270,width=\columnwidth]{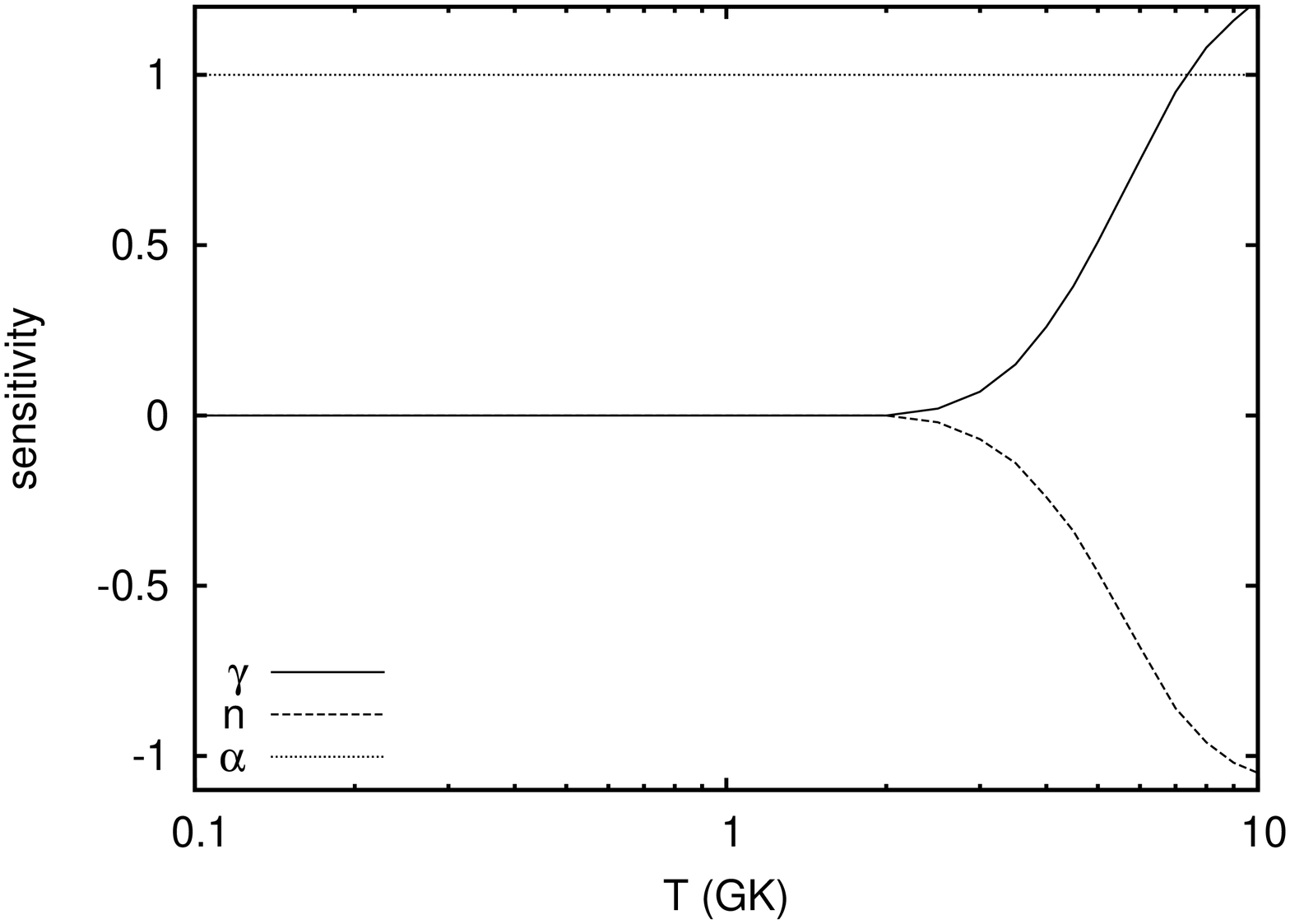}
\caption{\label{fig:agratesensi} Sensitivity of the $^{127}$I($\alpha$,$\gamma$)$^{131}$Cs stellar reaction rate to a variation of the $\gamma$, neutron, and $\alpha$ width as function of plasma temperature $T$. The sensitivity to a variation of the $\alpha$ width
is constantly unity at all temperatures. The temperature relevant for this reaction in the astrophysical $\gamma$ process is $2\leq T\leq 2.5$ GK.}
\end{figure}

Figure \ref{fig:ag} compares the standard NON-SMOKER prediction \cite{adndt1} (dashed line) to the ($\alpha$,$\gamma$) data. The calculation overestimates the experimental $S$ factor by a factor of two at the upper end of the investigated energy range and a factor of about 6 at the lower end. Also shown are calculations with the code SMARAGD \cite{SMARAGD}, version 0.8.4s, using different numbers of discrete excited states (this issue is discussed below in Sec.~\ref{sec:exc}). The SMARAGD calculation with 5 discrete states included is similar to the NON-SMOKER values, although slightly closer to the data. As pointed out in Sec.~\ref{sec:intro}, however, one should not jump to conclusions regarding the prediction of the stellar $\alpha$ capture rate. The sensitivity of the stellar reaction rate to a variation of the averaged $\gamma$, neutron, and $\alpha$ widths by a factor of two is shown in Fig.~\ref{fig:agratesensi}. (All discussed cross sections and rates are insensitive to a variation of the proton width, therefore it is never shown.) A sensitivity value $|s|=1$ implies that the change in the cross section or rate is of the same magnitude as the variation factor, a value $s=0$ shows that there is no dependence on this width \cite{sensi}. The sign of the plotted sensitivities informs whether the cross section or rate changes in the same direction as or opposite to the width. We note that the ($\alpha$,$\gamma$) cross sections and rates change inversely proportional to the neutron width whereas they depend proportionally on the $\gamma$ and/or $\alpha$ width. The dependences are reversed between neutron and $\gamma$ width in the ($\alpha$,n) channel. For a detailed explanation of the possible cases, see \cite{sensi}.

\begin{figure}
\includegraphics[angle=270,width=\columnwidth]{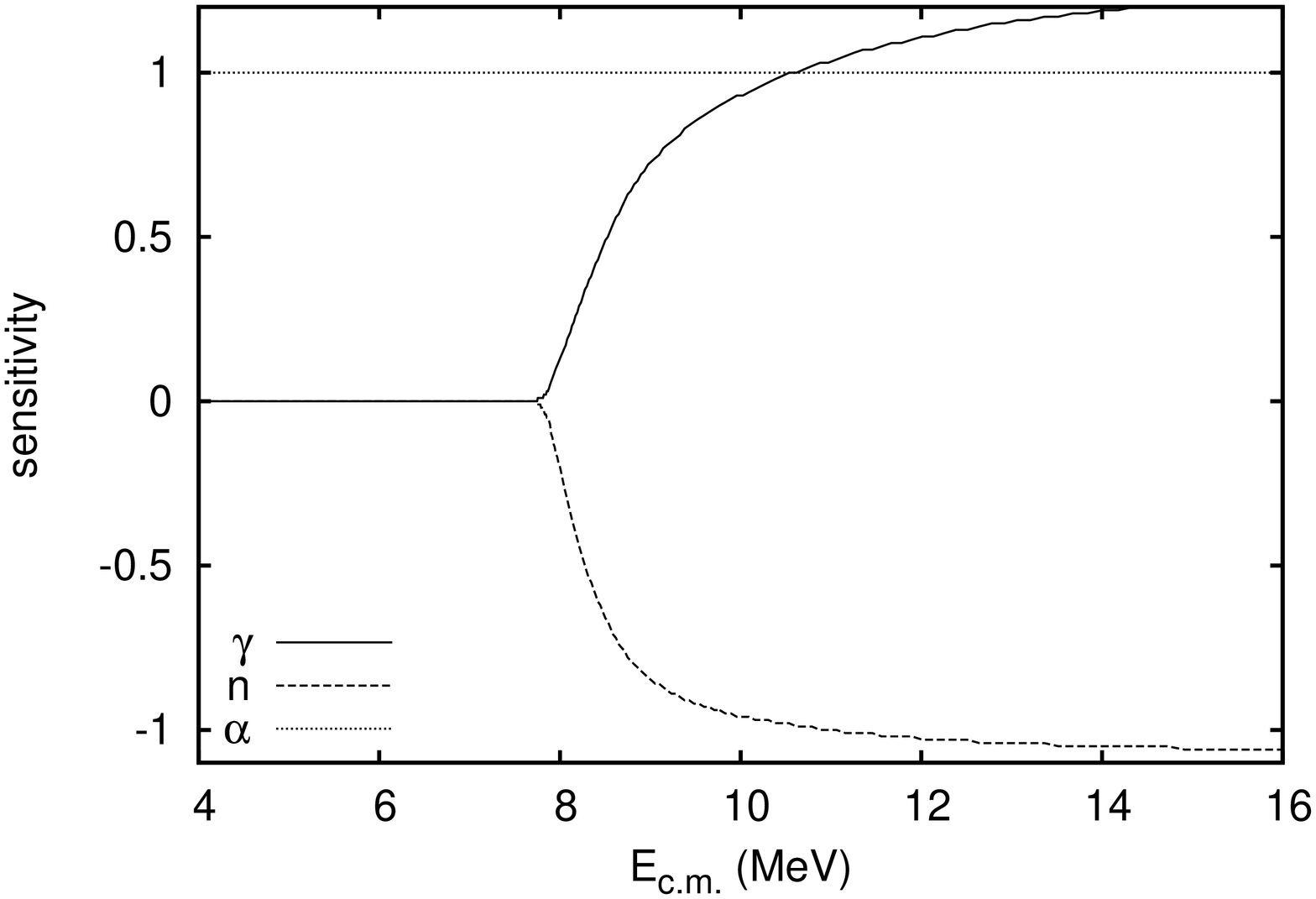}
\caption{\label{fig:agsensi} Sensitivity of the $^{127}$I($\alpha$,$\gamma$)$^{131}$Cs reaction cross sections to a variation of the $\gamma$, neutron, and $\alpha$ width as function of c.m.\ energy. The sensitivity to a variation of the $\alpha$ width
is constantly unity at all energies.}
\end{figure}

As can be seen in Fig.~\ref{fig:agratesensi}, the reaction rate is only sensitive to the $\alpha$ width at the temperature region relevant in the $\gamma$ process. Comparing this to the sensitivity of the capture cross sections in Fig.~\ref{fig:agsensi} it is immediately obvious that, in addition to the $\alpha$ widths, those also strongly depend on the $\gamma$ and neutron widths in the measured energy range. Only below the ($\alpha$,n) threshold, the situation is similar to the one of the rate. Therefore it is impossible to further disentangle the different contributions of the widths to the total deviation from experiment and to draw strong conclusions on the basis of the experimental capture data alone.

\begin{figure}
\includegraphics[angle=270,width=\columnwidth]{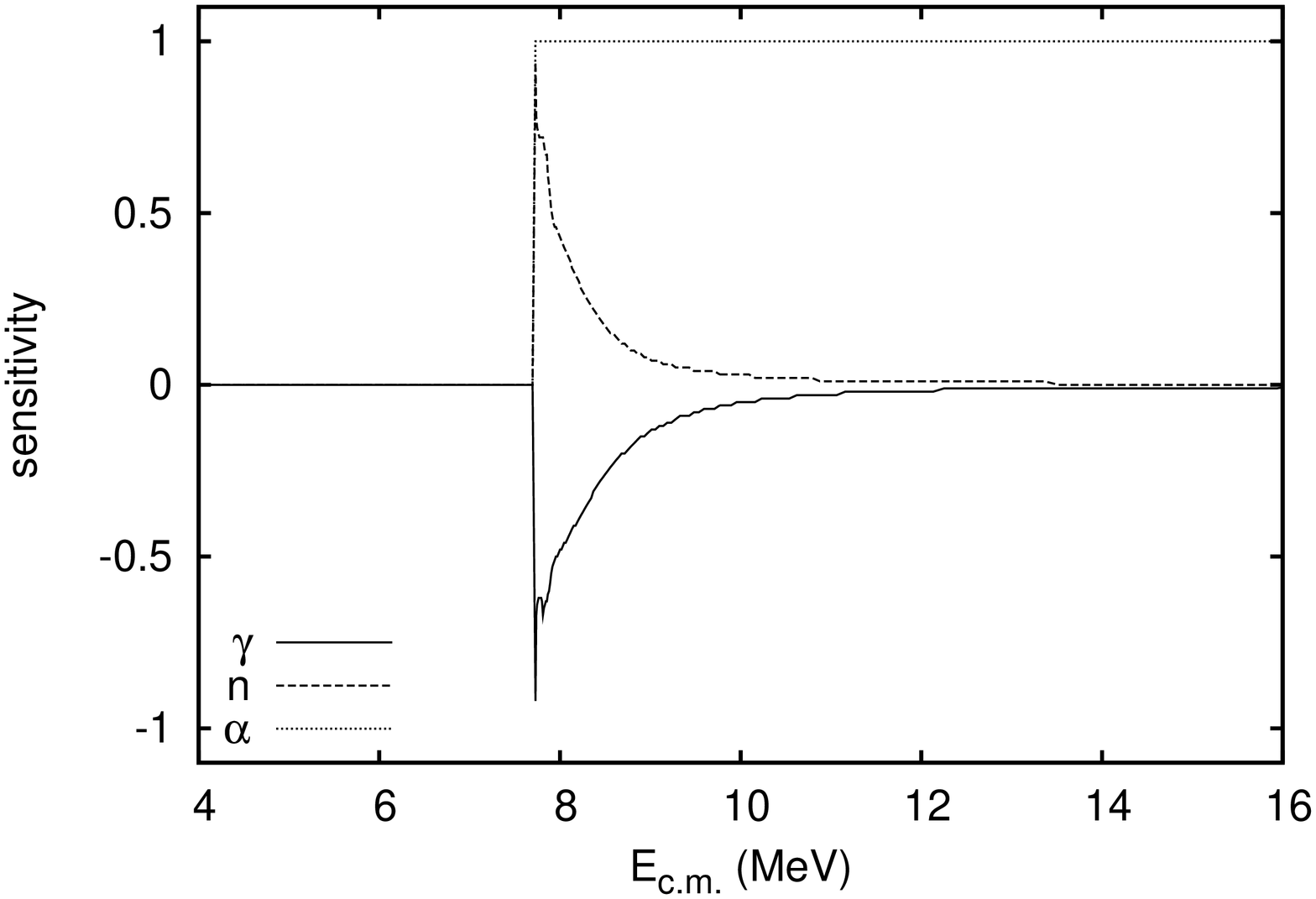}
\caption{\label{fig:ansensi} Sensitivity of the $^{127}$I($\alpha$,n)$^{130}$Cs reaction cross sections to a variation of the $\gamma$, neutron, and $\alpha$ width as function of c.m.\ energy. The sensitivity to a variation of the $\alpha$ width
is constantly unity at all energies above the threshold.}
\end{figure}

The ($\alpha$,n) cross sections (and $S$ factors) at the upper end of the measured energy range, on the other hand, also are only sensitive to the $\alpha$ width, similar to the reaction rate, as seen in Fig.~\ref{fig:ansensi}. Inspection of Fig.~\ref{fig:an}, comparing the ($\alpha$,n) data to the calculations, finds that the theory values are close to the experimental ones, especially at higher energies. While NON-SMOKER is slightly above the data, the SMARAGD calculation using five excited states is in good agreement with the measured $S$ factors. It is to be concluded, therefore, that the $\alpha$ width is predicted well by the SMARAGD code and any deviations from the ($\alpha$,$\gamma$) experiment are due to problems in the neutron and/or $\gamma$ widths. Without further information, it is impossible to further identify one of these widths as main source of error.

Since the g.s.\ contribution to the stellar rate is $X=0.35$ at 2.5 GK \cite{sensi}, the experimental data cannot be used to compute the \textit{stellar} capture and photodisintegration rates. The $\alpha$ width, however, was shown to be the important width for the rate and its prediction was confirmed by the data, at least in the measured energy range. Therefore the stellar rate can be calculated utilizing the same approach that was used for comparison with the data but including the thermal plasma effects. The SMARAGD code uses a superior method to calculate transmission coefficients and averaged widths below the Coulomb barrier compared to the NON-SMOKER code \cite{raureview} and we used this code (with five discrete excited states included) to provide the stellar reactivities in Table \ref{tab:rates}. The REACLIB parameters \cite{adndt} for a fit based on the new reactivity values are given in Table \ref{tab:pars}. The fit accuracy is better than 0.2\% within the relevant temperature range.

\begin{figure}
\includegraphics[angle=270,width=\columnwidth]{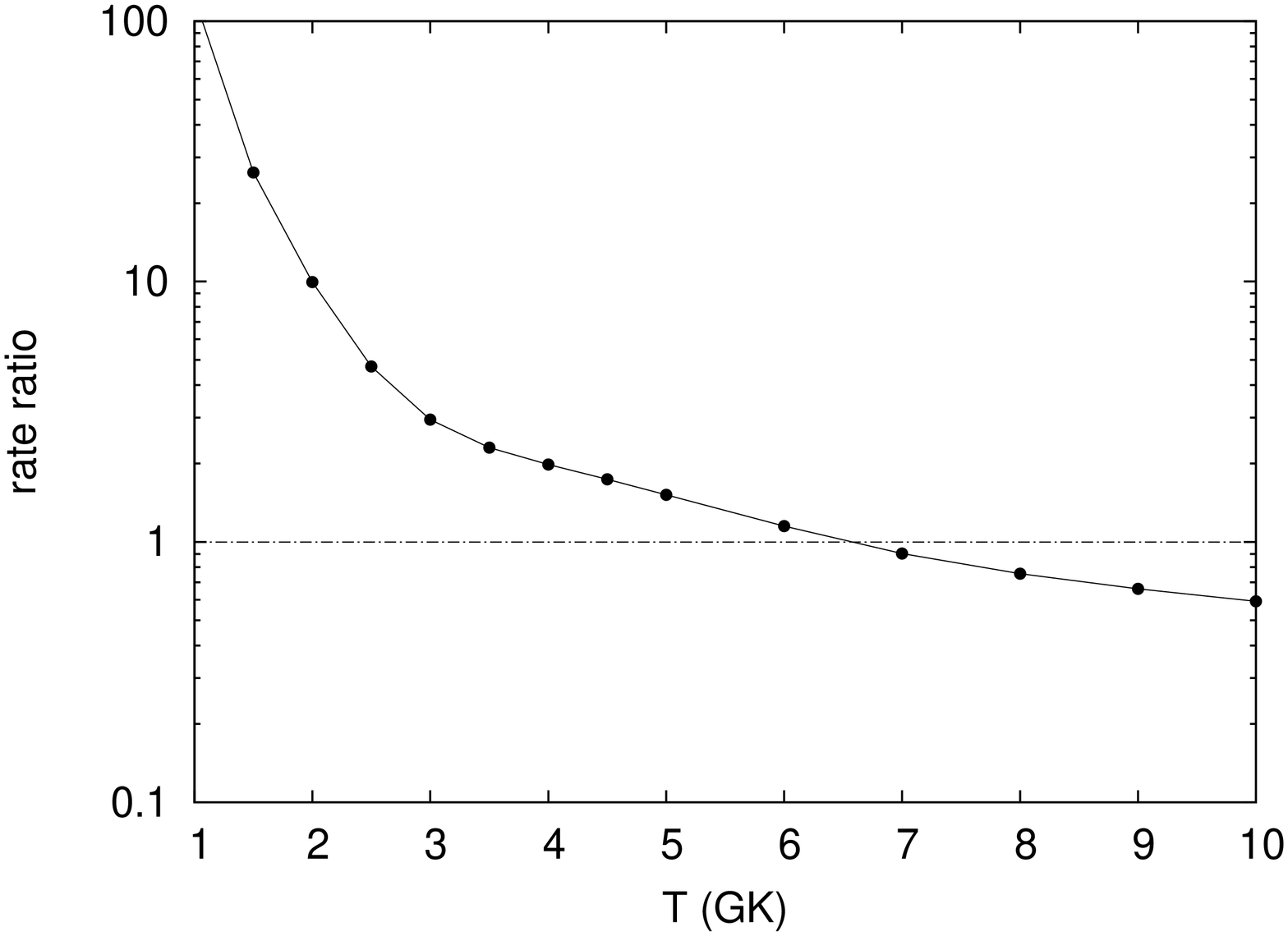}
\caption{\label{fig:rateratio} Ratio SMARAGD/NON-SMOKER of the newly calculated SMARAGD stellar reaction rate for $^{127}$I($\alpha$,$\gamma$)$^{131}$Cs and the NON-SMOKER predictions \cite{adndt1}. The horizontal line at unity ratio is drawn to guide the eye.}
\end{figure}

Figure \ref{fig:rateratio} shows a comparison of the previous standard rate from \cite{adndt1} (which was fitted to provide the REACLIB parameters in \cite{adndt}). At 2.0 and 2.5 GK, the rate is increased by factors $10$ and $4$, respectively. This is neither a consequence of different optical $\alpha$+nucleus potentials (since both codes use the same \cite{mcf}) nor of a modification based on the new experimental data. It simply shows the difference in the numerical solution of the Schr\"odinger equation at strongly subCoulomb energies \cite{raureview}.

\begin{table}
\caption{\label{tab:rates}Stellar reactivity $N_A \left< \sigma v \right>^*$ for $^{127}$I($\alpha$,$\gamma$)$^{131}$Cs as function of plasma temperature $T$.}
\begin{ruledtabular}
\begin{tabular}{rl}
\multicolumn{1}{c}{$T$} & \multicolumn{1}{c}{Reactivity} \\
\multicolumn{1}{c}{(GK)} & \multicolumn{1}{c}{(cm$^3$s$^{-1}$mole$^{-1}$)} \\
\hline
  0.50 & $  1.32\times 10^{-41} $ \\
  0.60 & $  3.79\times 10^{-37} $ \\
  0.70 & $  1.26\times 10^{-33} $ \\
  0.80 & $  9.34\times 10^{-31} $ \\
  0.90 & $  2.36\times 10^{-28} $ \\
  1.00 & $  2.64\times 10^{-26} $ \\
  1.50 & $  3.04\times 10^{-19} $ \\
  2.00 & $  5.83\times 10^{-15} $ \\
  2.50 & $  5.00\times 10^{-12} $ \\
  3.00 & $  6.69\times 10^{-10} $ \\
  3.50 & $  2.58\times 10^{-8} $ \\
  4.00 & $  4.18\times 10^{-7} $ \\
  4.50 & $  3.66\times 10^{-6} $ \\
  5.00 & $  2.03\times 10^{-5} $ \\
  6.00 & $  2.51\times 10^{-4} $ \\
  7.00 & $  1.41\times 10^{-3} $ \\
  8.00 & $  4.86\times 10^{-3} $ \\
  9.00 & $  1.19\times 10^{-2} $ \\
 10.00 & $  2.23\times 10^{-2} $
\end{tabular}
\end{ruledtabular}
\end{table}

\begin{table}
\caption{\label{tab:pars}REACLIB parameters \cite{adndt} obtained from fitting the reactivities shown in Table \ref{tab:rates}.}
\begin{ruledtabular}
\begin{tabular}{ccc}
Parameter & \multicolumn{1}{c}{($\alpha$,$\gamma$)} & \multicolumn{1}{c}{($\gamma$,$\alpha$)} \\
\hline
$a_0$  & $-2.092540\times 10^{3}$ & $-2.066396\times 10^{3}$ \\
$a_1$  & $-4.266646\times 10^{2}$ & $-4.440888\times 10^{2}$  \\
$a_2$  & \multicolumn{2}{c}{$7.026704\times 10^{3}$}  \\
$a_3$  &  \multicolumn{2}{c}{$-4.701040\times 10^{3}$}   \\
$a_4$  & \multicolumn{2}{c}{$1.388502\times 10^{2}$}  \\
$a_5$  & \multicolumn{2}{c}{$-4.682267$}  \\
$a_6$  & $3.399136\times 10^{3}$ & $3.400636\times 10^{3}$
\end{tabular}
\end{ruledtabular}
\end{table}

Finally, it has to be cautioned that the $\alpha$ width is still not well constrained at astrophysically relevant energies by the experiment. It is conceivable, for instance, that the subCoulomb energy dependence is different than the one found in the measured energy range. In fact, there could be an indication of this at the lowest two data points. Inspecting Fig.~\ref{fig:an}, there seems to be a slight deviation below 11 MeV. One might be tempted to assign this to the same problem with the neutron and/or $\gamma$ width showing up in the capture data because it appears just as the sensitivity to the neutron and $\gamma$ widths increases. Comparing the sensitivities of the ($\alpha$,$\gamma$) and ($\alpha$,n) reaction, however, reveals that the required modification would be different in the two cases. The calculated ($\alpha$,$\gamma$) excitation function could be shifted to lower values by either increasing the $\gamma$ width or decreasing the neutron width. Both changes would act oppositely in the ($\alpha$,n) excitation function, moving the theory values further away from the data at low energy. Therefore, if this is the sign of a real effect, it has to be due to a changed energy dependence in the $\alpha$ width. The indication for this in the current data, however, is too vague to draw any conclusion.

\subsection{The problem of the selection of discrete excited states in Hauser-Feshbach calculations}
\label{sec:exc}

These reactions are also well suited to discuss another problem appearing in the prediction of reactions, especially for unstable nuclei close to stability with partially known level schemes. The H-F transmission coefficients, required to compute the total width in each reaction channel, are calculated from sums of individual transitions from a compound nuclear state to all energetically possible final states. In each reaction channel excited states have to be considered up to an energy $E_\mathrm{max}=E_\mathrm{c.m.}+Q_\mathrm{c}$. Depending on the reaction $Q$ value of the channel $Q_\mathrm{c}$, often discrete excited levels are only known up to an energy $E_\mathrm{last}<E_\mathrm{max}$. Then the calculation of transitions to levels above $E_\mathrm{last}$ are performed by an integration over a nuclear level density instead of a sum over discrete, isolated states \cite{raureview,sensi}.

Far off stability, only few or no discrete states are known and the level density integration is used from low excitation energies onwards or directly above the g.s. But also close to or at stability, where levels are often specified up to $E_\mathrm{max}$, it is necessary to choose an appropriate cut-off at an energy below $E_\mathrm{max}$ because it is essential for a successful prediction to have a \textit{complete} level scheme included. Often, even when levels are given up to high excitation energy in a database like NuDAT or ENSDF \cite{nudat,ENSDF}, these are only partial level schemes, derived from a limited number of experiments or coming from information on rotational bands. At higher excitation energies it becomes also harder to experimentally resolve individual levels lying close to each other. An educated guess has to be made when preparing the input for H-F calculations, estimating up to which excitation energy the experimental level scheme available in databases can be considered complete. This is especially important in particle channels because particle transitions to low-lying excited states contribute most, whereas in the $\gamma$ channel transitions to higher excitation energies (with lower relative $\gamma$ energy) are dominant \cite{sensi,gammaenergies} and lead to the region where the nuclear level density is used, anyway.

\begin{table}
\caption{\label{tab:excited}Excited states in $^{130}$Cs used in the H-F calculations.}
\begin{ruledtabular}
\begin{tabular}{rrrr}
\multicolumn{2}{c}{NON-SMOKER \cite{adndt1,adndt}} & \multicolumn{2}{c}{SMARAGD \cite{raureview,SMARAGD}} \\
\multicolumn{1}{c}{Energy} & $J^\pi$ & \multicolumn{1}{c}{Energy} & $J^\pi$ \\
\multicolumn{1}{c}{(MeV)} &  & \multicolumn{1}{c}{(MeV)} & \\
\hline
0.0000 & 1$^+$ &    0.0000 & 1$^+$ \\
0.0804 & 2$^+$ &    0.0804 & 2$^+$ \\
0.1315 & 2$^+$ &    0.1315 & 2$^+$ \\
0.1484 & 2$^-$ &    0.1484 & 2$^-$ \\
0.1633 & 5$^-$ &    0.1633 & 5$^-$ \\
\hline
       &       &    0.2788 & 6$^-$ \\
       &       &    0.3756 & 7$^-$ \\
       &       &    0.5655 & 8$^-$ \\
       &       &    0.6181 & 8$^-$ \\
       &       &    0.8783 & 9$^-$ \\
       &       &    0.9748 & 9$^+$ \\
       &       &    0.9972 & 9$^-$ \\
       &       &    1.1265 & 10$^+$ \\
       &       &    1.1721 & 10$^-$ \\
       &       &    1.2654 & 10$^-$ \\
       &       &    1.4798 & 11$^+$ \\
       &       &    1.5129 & 11$^-$ \\
       &       &    1.6738 & 11$^+$ \\
       &       &    1.7700 & 12$^+$ \\
       &       &    1.9607 & 12$^-$ \\
       &       &    2.0748 & 12$^+$ \\
       &       &    2.0861 & 12$^-$ \\
       &       &    2.1870 & 13$^+$ \\
       &       &    2.4468 & 13$^+$ \\
       &       &    2.6135 & 14$^+$ \\
       &       &    2.7966 & 14$^+$ \\
       &       &    3.0825 & 15$^+$ \\
       &       &    3.2498 & 15$^+$ \\
       &       &    3.5475 & 16$^+$ \\
       &       &    4.0405 & 17$^+$
\end{tabular}
\end{ruledtabular}
\end{table}

The $Q$ value of the reaction $^{127}$I($\alpha$,n)$^{130}$Cs is $-7.729$ MeV and therefore the impact of the low-lying excited states is even more pronounced.
Table \ref{tab:excited} lists the set of discrete excited states in $^{130}$Cs used in the H-F calculations shown in Figs.~\ref{fig:ag} and \ref{fig:an}. 
Only five discrete levels, including the g.s., were included in the NON-SMOKER calculation, as given in \cite{adndt1}. Above the fourth excited state, a theoretical nuclear level density \cite{rtk} was used. Version 0.8.4s of the SMARAGD code made use of an updated level scheme from \cite{ENSDF}, also included in the 2010 version of NuDAT \cite{nudat}. Above the last included discrete state, a refitted version of the level density of \cite{rtk} is applied and additionally an energy-dependent parity distribution \cite{darko}.

Using the full list of 30 levels given up to 4.04 MeV, the curve marked "SMARAGD(30 exc.\ stat.)" in Figs.~\ref{fig:ag}, \ref{fig:an} is obtained. Since $^{127}$I($\alpha$,$\gamma$)$^{131}$Cs is also sensitive to the neutron width (see Fig.~\ref{fig:agsensi}), it is also affected. The artificial break at about 11.8 MeV originates from the inclusion of transitions calculated with a theoretical nuclear level density above this energy. It is a clear sign that the number of included discrete levels close to this energy is quite different from the predicted number. This is due to an incomplete level scheme. 
As can be seen in Table \ref{tab:excited} and in Ref.~\cite{ENSDF}, the given states at higher energy are only states with higher spin from four rotational bands, low-spin states are not identified. Truncating the list of discrete levels after the fifth one yields the curve labeled "SMARAGD(5 exc.\ stat.)" in Figs.~\ref{fig:ag}, \ref{fig:an}. It is to be noted that the choice of levels obviously has no impact below the ($\alpha$,n) threshold and therefore there is also no impact on the astrophysical rate for the ($\alpha$,$\gamma$) reaction.

\section{Summary and conclusions}
\label{sec:sum}

The reactions $^{127}$I($\alpha$,$\gamma$)$^{131}$Cs and $^{127}$I($\alpha$,n)$^{130}$Cs have been studied using the activation technique combining X- and $\gamma$-ray countings. For the first time, measurements have been performed for these reactions close to the astrophysically relevant energy region, at energies  $9.50\leq E_{c.m.} \leq 15.15$ MeV and  $9.62\leq E_{c.m.} \leq 15.15$ MeV, respectively. Furthermore, the relative intensity of the 536.1 keV $\gamma$ transition was measured precisely, its uncertainty is reduced from 13\% to 4\%.

The ($\alpha$,n) data confirm the H-F predictions of the averaged $\alpha$ width which is essential to derive the astrophysical $^{127}$I($\alpha$,$\gamma$) and $^{131}$Cs($\gamma$,$\alpha$) rates, important in the nucleosynthesis of heavy $p$ nuclei in a $\gamma$ process. The ($\alpha$,$\gamma$) data reveal deficiencies in the prediction of the $\gamma$ and/or neutron width within the measured energy range. This is inconsequential, however, for the stellar ($\alpha$,$\gamma$) and ($\gamma$,$\alpha$) rates. Despite an unchanged optical $\alpha$+nucleus potential, a recalculation of the reaction rates with an improved code yielded increased rates compared to previous calculations at $\gamma$ process temperatures.

There is an indication of an onset of change in the energy dependence of the $\alpha$ width in the ($\alpha$,n) data at the two lowest energies. It would be ironic if there is a reduction of the $\alpha$ width at very low $\alpha$ energies bringing the new rate down to the previous value. The current data, however, is not sufficient to arrive at a firm conclusion. More data at lower energies, preferably below the ($\alpha$,n) threshold, would be required to clarify this issue.

\begin{acknowledgments}
This work was supported by the EUROGENESIS research
program, the European Research Council starting Grant no. 203175,  OTKA (NN83261, K101328, PD104664), Tubitak (108T508 No. 109T585) and the ENSAR/THEXO European FP7 programme. G. G. Kiss acknowledges support from the Bolyai grant and Z. Korkulu acknowledges support from the ERASMUS Programme of the European Commission.
This work was also supported by the T\'AMOP-4.2.2/B-10/1-2010-0024 project.
\end{acknowledgments}

\end{document}